\documentclass[twocolumn,superscriptaddress,amsmath,amssymb,floats]{revtex4-1}
\usepackage[latin9]{inputenc}
\usepackage{color}
\usepackage{mathtools}
\usepackage{amsmath}
\usepackage{amssymb}
\usepackage{graphicx}
\usepackage{verbatim}
\usepackage{epstopdf}
\usepackage{comment}
\usepackage{tikz}

\makeatletter

\@ifundefined{textcolor}{}
{%
 \definecolor{BLACK}{gray}{0}
 \definecolor{WHITE}{gray}{1}
 \definecolor{RED}{rgb}{1,0,0}
 \definecolor{GREEN}{rgb}{0,1,0}
 \definecolor{BLUE}{rgb}{0,0,1}
 \definecolor{CYAN}{cmyk}{1,0,0,0}
 \definecolor{MAGENTA}{cmyk}{0,1,0,0}
 \definecolor{YELLOW}{cmyk}{0,0,1,0}
}

\usepackage{epstopdf}
\usepackage{array}
\usepackage{mathrsfs}
\usepackage{dcolumn}
\usepackage{bm}

\newcolumntype{L}[1]{>{\raggedright\let\newline\\\arraybackslash\hspace{0pt}}m{#1}}
\newcolumntype{C}[1]{>{\centering\let\newline\\\arraybackslash\hspace{0pt}}m{#1}}
\newcolumntype{R}[1]{>{\raggedleft\let\newline\\\arraybackslash\hspace{0pt}}m{#1}}

\newcommand{\mbf}[1]{\mathbf{#1}}

\usepackage{ulem}

\usepackage{tikz}
\usetikzlibrary{calc,intersections}
\usetikzlibrary{shadings}
\usepgflibrary{shadings}

\makeatother

\begin{document}

\title{Role of multiorbital effects in the magnetic phase diagram of iron pnictides}

\author{Morten H. Christensen}

\affiliation{Niels Bohr Institute, University of Copenhagen, Juliane Maries Vej 30, DK-2100, Denmark}

\author{Daniel D. Scherer}

\affiliation{Niels Bohr Institute, University of Copenhagen, Juliane Maries Vej 30, DK-2100, Denmark}

\author{Panagiotis Kotetes}

\affiliation{Center for Quantum Devices, Niels Bohr Institute, University of Copenhagen, Universitetsparken 5, DK-2100, Denmark}

\author{Brian M. Andersen}

\affiliation{Niels Bohr Institute, University of Copenhagen, Juliane Maries Vej 30, DK-2100, Denmark}

\begin{abstract}

We elucidate the pivotal role of the bandstructure's orbital content in deciding the type of commensurate magnetic order stabilized within the itinerant scenario of iron-pnictides. Recent experimental findings in the tetragonal magnetic phase attest to the existence of the so-called charge and spin ordered density wave over the spin-vortex crystal phase, the latter of which tends to be favored in simplified band models of itinerant magnetism. Here we show that employing a multiorbital itinerant Landau approach based on realistic bandstructures can account for the experimentally observed magnetic phase, and thus shed light on the importance of the orbital content in deciding the magnetic order. In addition, we remark that the presence of a hole pocket centered at the Brillouin zone's ${\rm M}$-point favors a magnetic stripe rather than a tetragonal magnetic phase. For inferring the symmetry properties of the different magnetic phases, we formulate our theory in terms of magnetic order parameters transforming according to irreducible representations of the ensuing D$_{\rm 4h}$ point group. The latter method not only provides transparent understan\-ding of the symmetry breaking schemes but also reveals that the leading instabilities always belong to the $\{A_{1g},B_{1g}\}$ subset of irreducible representations, independent of their C$_2$ or C$_4$ nature. 

\end{abstract}

\maketitle

\section{introduction}

The importance of a nested bandstructure and the concomitant emergence of itinerant magnetism have recently come to the fore in the field of iron-based superconductors (FeSC). A detailed understanding of the microscopic mechanism driving magnetism in these systems is signi\-fi\-cant and interesting in its own right, but can also lead to new insights regarding the emergence of superconducti\-vi\-ty upon hole or electron doping~\cite{scalapinormp}. At present, a conclusive microscopic description of the origin of the various magnetic phases of the FeSC along with their evolution with temperature, doping, and pressure remains an open problem. 

The prevalent magnetic phase of the FeSC is characte\-ri\-zed by a metallic spin-density wave (SDW) with orde\-ring wavevectors $\mbf{Q}_{X}=(\pi,0)$ or $\mbf{Q}_{Y}=(0,\pi)$, and is also known as the magnetic stripe (MS) phase. In the latter, magnetism is collinear and additio\-nal\-ly introduces an orthorhombic distortion to the otherwise C$_4$-symmetric Fe crystal lattice. In addition, the magnetic moments are oriented ferromag\-neti\-cal\-ly (antiferromagnetically) along the shorter (longer) axis. However, early theoretical studies pointed out the potential relevance of additional so-called double-$\mbf{Q}$ magnetic phases, in which simultaneous ordering at both $\mbf{Q}_{X,Y}$ wavevectors takes place~\cite{lorenzana08,eremin,brydon,giovannetti}. There are two distinct double-$\mbf{Q}$ magnetic phases allowed to appear at a paramagnetic(PM)-magnetic transition: a collinear nonuniform charge- and spin-ordered density wave (CSDW) phase, and a non-collinear so-called spin-vortex crystal (SVC) phase where the moments on neighboring sites are at right angles to each other (see Fig.~\ref{fig:commensurate_phase_diagram}, top panel). 

So far, a series of theoretical studies have explored the phase diagram, and determined the preferred magnetic state, MS, CSDW or SVC, in terms of temperature, electron filling, microscopic interaction paramenters, spin-orbit coupling (SOC), and disorder~\cite{lorenzana08,eremin,brydon,giovannetti,fernandes12,gastiasoro15,kang15a,christensen15,hoyer16a,Ohalloran,wang17,gastiasoro17}. Nevertheless, it is still unclear which features of a given bandstructure are responsible for selecting the magnetic state, or how this choice depends on the values of the interactions. Detailed theoretical calculations are thus required in each seperate case to accurately determine the global minimum of the magnetic free energy. On top of that, a full consensus on the phase diagram is lacking; whereas free energy expansions in terms of the magnetic order pa\-ra\-me\-ter in three-band models consistently find SVC and MS phases~\cite{fernandes12,kang15a,wang15,hoyer16a}, microscopic Hartree-Fock stu\-dies at low temperatures show the existence of both SVC, MS, and CSDW phases in the typical tempe\-ra\-tu\-re versus do\-ping phase diagram~\cite{gastiasoro15,scherer16}.

Experimentally, the presence of double-$\mbf{Q}$ magnetism has been confirmed via employing diverse techniques in a number of FeSC materials, including hole-doped Ba$_{1-x}$Na$_x$Fe$_2$As$_2$~\cite{avci14a,wang16}, Ba$_{1-x}$K$_x$Fe$_2$As$_2$~\cite{hassinger,bohmer15a,allred15a,malletta,mallettb}, Sr$_{1-x}$Na$_x$Fe$_2$As$_2$~\cite{allred16a,taddei16}, and most recently Ca$_{1-x}$Na$_x$Fe$_2$As$_2$~\cite{taddei17}. Notably, scattering methods have proven suitable for differentiating between the C$_4$ and C$_2$ phases, by comparison of magnetic and crystal Bragg peaks. The C$_4$-symmetric phases have been so far detected at the foot of the magnetic dome consi\-sting mainly of MS orde\-ring, in agreement with some theoretical studies~\cite{gastiasoro15}. Recently, the site-sensitivity of M\"{o}ssbauer spectroscopy was exploited to conclude that in Sr$_{1-x}$Na$_x$Fe$_2$As$_2$, the relevant magnetic double-$\mbf{Q}$ ordered phase exhibits only moments on half of the Fe ions~\cite{allred16a}, in agreement with the CSDW phase, and the general itinerant scenario for magnetism in these compounds~\cite{chubukov}.

Fuelled by these continued experimental findings and the above-mentioned theoretical puzzle, we show here that accounting for the multiorbital character of the bandstructure in the free energy expansion yields improved agreement with the experimentally inferred magnetic phases compared to models that neglect the orbital structure of the band. In more detail, we retrieve the commensurate magnetic phase diagram upon varying the profile of the bandstructure, that here is based on density functional theory (DFT) studies~\cite{ikeda10}. Using the Landau formalism, we identify the leading magnetic instability and evaluate the relevant Landau coefficients in order to determine the system's preferred magnetic state and the symmetry of the respective magnetic order parameter.
Here we focus on commensurate magnetic phases, for incommensurate ordering vectors the expression for the Landau free energy is modified, and new magnetic phases arise as shown recently~\cite{christensen16}. Within our approach the orbital content of the bands can be excluded when evaluating the ground state magnetic order. This allows the direct comparison of our results to the ones obtained via previous band mo\-dels that neglected the orbital content.

In addition, we pursue resolving the influence of the hole pocket centered at the ${\rm M}$-point, on the phase diagram, as it can be present or absent depending on the FeSC material under investigation. Our results suggest that the presence of the above hole pocket tends to favor the emergence of the MS phase at the expense of the other two C$_4$-symmetric phases. Finally we note that a crucial component of our theoretical analysis is to classify the magnetic order parameters according to the irreducible representations (IRs) of the D$_{\rm 4h}$ point group that cha\-rac\-te\-ri\-zes the bandstructure under consideration. In this manner we find that the order parameters of the leading instabi\-li\-ties belong to particular IR subsets, irrespective of the C$_2$ or C$_4$ nature of the magnetic ground state. 

Below we provide a summary of our main results, that we discuss in more detail in the upcoming sections.

\begin{itemize}

\item The magnetic order parameters belong to the following three different subsets $\{A_{1g},B_{1g}\}$, $\{A_{2g},B_{2g}\}$ and $E_{1g}$, formed by IRs of D$_{\rm 4h}$. The magnetic order parameters with IRs of a given subset share the same critical temperature. For the cases studied here the magnetic order parameter of the leading instability was always found to belong to the $\{A_{1g},B_{1g}\}$ IR subset. This implies a nearly dia\-go\-nal magnetic order parameter matrix in orbital space, with off-diagonal components of $\{x^2-y^2,z^2\}$ character, as highlighted in Eq.~(\ref{eq:A1g_B1g_structure}). We stress that the form of the order parameters presented in Eqs.~(\ref{eq:A1g_B1g_structure})--(\ref{eq:Eg_structure}) is not a property of any specific bandstructure but a generic consequence of the point group symmetries.

\item Excluding the orbital content promotes the SVC phase over the CSDW phase in agreement with previous theoretical calculations~\cite{wang15,hoyer16a}. Instead, ta\-king into account the orbital degrees of freedom at or near the FS, captures to a large extent the salient features of the case with all excitations in the Fe-3$d$ shell included. For more details see Fig.~\ref{fig:phase_diagrams_U085}.

\item For the bandstructure studied here, we find that the M-pocket plays an important role in selecting the single-$\mbf{Q}$ over the double-$\mbf{Q}$ magnetic phases, as shown in Fig.~\ref{fig:m_pocket_phase_diagram}. In the absence of a hole-pocket at M, the value of the quartic coefficient deciding between C$_2$ and C$_4$ phases is reduced, thus favoring a C$_4$ phase. Nevertheless, the C$_2$ phase is re\-co\-ve\-red at lower temperatures and larger interaction strengths.

\end{itemize}
All our results from the Landau approach are compared one-to-one with results from Hartree-Fock calculations. As expected, the two methods yield identical results close to the transition temperature $T_{\rm SDW}$ of the magnetic phase. Our results thus corroborate the importance of accounting for the full bandstructure when modelling the magnetic phase diagrams of FeSC.

\section{microscopic model and IR decomposition }

To study commensurate magnetism within the framework of itinerant electrons we adopt the multiorbital Hubbard-Hund Hamiltonian
\begin{eqnarray}
	\mathcal{H} &=& \mathcal{H}_0 + \mathcal{H}_{\rm int}\,,\\
	\mathcal{H}_0 &=& \sum_{\mbf{k}} \sum_{\substack{ab \\ \sigma}}\left(\epsilon_{ab}(\mbf{k}) - \mu \delta_{ab} \right)c^{\dagger}_{\mbf{k} a \sigma} c^{\phantom{\dagger}}_{\mbf{k}b\sigma}\,, \\
	\mathcal{H}_{\rm int} &=& \frac{U}{\mathcal{N}}\sum_{\mbf{q}}\sum_{a}n_{\mbf{q}a\uparrow}n_{\mbf{-q}a\downarrow} + \frac{U'}{\mathcal{N}}\sum_{\mbf{q}}\sum_{\substack{a < b \\ \sigma\sigma'}}n_{\mbf{q}a\sigma}n_{\mbf{-q}b\sigma'} \nonumber \\ && + \frac{J}{2\mathcal{N}}\sum_{\mbf{k}\mbf{k}^{\prime}\mbf{q}}\sum_{\substack{a \neq b \\ \sigma\sigma'}}c_{\mbf{k+q}a\sigma}^{\dagger}c_{\mbf{k}b\sigma}c_{\mbf{k'-q}b\sigma'}^{\dagger}c_{\mbf{k'}a\sigma'} \nonumber \\ && + \frac{J'}{2\mathcal{N}}\sum_{\mbf{k}\mbf{k}^{\prime}\mbf{q}}\sum_{\substack{a\neq b \\ \sigma}}c_{\mbf{k+q}a\sigma}^{\dagger}c_{\mbf{k'-q}a\bar{\sigma}}^{\dagger}c_{\mbf{k'}b\bar{\sigma}}c_{\mbf{k}b\sigma}\,,\label{eq:multi_orbital_interaction}
\end{eqnarray}
where the momentum sums run over the 1-Fe Brillouin zone (BZ), $\mathcal{N}$ being the number of momentum space points, the $\epsilon_{ab}(\mbf{k})$ are taken from Table II, upper right, in Ref.~\cite{ikeda10} (LaFeAsO) and $n_{\mbf{q}a\sigma} = \sum_{\mbf{k}}c^{\dagger}_{\mbf{k}+\mbf{q}a\sigma}c^{\phantom{\dagger}}_{\mbf{k}a\sigma}$. Here $a$ and $b$ label Fe 3$d$-orbitals $ \{ xz,\,{yz},\,{xy},\,{x^{2}-y^{2}},\,{z^{2}} \} $ and $\sigma$ labels spin, with $\bar{\sigma}=-\sigma$. We treat the interactions on a mean-field level and perform a Hartree-Fock decoupling in both the intraorbital $\mbf{q}=0$ charge channel and the $\mbf{q}=\mbf{Q}_{X,Y}=(\pi,0)/(0,\pi)$ magnetic channel. The $\mbf{q}=0$ mean-field densities are solved for in a selfconsistent manner resul\-ting in an orbitally dependent shift of the chemical potential $\tilde{\epsilon}_{a}$~\cite{commensurability}. The effects of doping are modelled by a rigid shift of the chemical potential $\mu$ thus leading to an appropriate alte\-ra\-tion of the electron filling.
The mean-field Hamiltonian is thus
\begin{eqnarray}
	\mathcal{H}^{\rm MF} &=& \sum_{\mbf{k}}\sum_{\substack{ab \\ \sigma}} \left( \epsilon_{ab}(\mbf{k}) - (\tilde{\epsilon}_{a}+\mu)\delta_{ab} \right) c^{\dagger}_{\mbf{k}a\sigma}c^{\phantom{\dagger}}_{\mbf{k}b\sigma}  \nonumber \\ 
	&& - \frac{1}{2}\sum_{\mbf{k}}\sum_{\substack{i \in \{ X, Y \}  \\ ab \\ \sigma\sigma'}} \left( \mbf{M}^{ab}_{i}\cdot c^{\dagger}_{\mbf{k} a\sigma} \boldsymbol{\sigma}_{\sigma\sigma'}c^{\phantom{\dagger}}_{\mbf{k}+\mbf{Q}_i b \sigma'} + \text{h.c.} \right) \nonumber \\
	&& + \mathcal{N}\sum_{\substack{i \in \{X,Y\} \\ abcd}} \left( U^{-1} \right)^{abcd} \mbf{M}^{ab}_{i} \cdot \mbf{M}^{cd}_{i}\,. \label{eq:mf_hamiltonian}
\end{eqnarray}
Here $\boldsymbol{\sigma}=\left(\sigma^{x} \ \sigma^{y} \ \sigma^{z} \right)^{T}$, $U^{abcd}$ correspond to the Hubbard-Hund interaction parameters
\begin{eqnarray}
	U^{aaaa} &=& U \qquad U^{abba} = U' \nonumber\\
	U^{abab} &=& J' \qquad U^{aabb} = J\,,
\end{eqnarray}
and
\begin{eqnarray}\label{eq:mag_mean_field_eq}
	\mbf{M}^{ab}_{i} = \frac{1}{2\mathcal{N}}\sum_{\mbf{k}}\sum_{\substack{cd \\ \sigma\sigma'}}U^{abcd}\langle c^{\dagger}_{\mbf{k}+\mbf{Q}_i c\sigma}\boldsymbol{\sigma}_{\sigma\sigma'}c^{\phantom{\dagger}}_{\mbf{k}d\sigma'} \rangle,\, i = X,\, Y.\,\,\quad
\end{eqnarray}
At this point we adopt the parameterization $U'=U-2J$ and $J'=J$, which is a consequence of imposing SO(3) rotational invariance on the sum of the interaction terms~\cite{oles83}.

Integrating out the electrons in Eq.~(\ref{eq:mf_hamiltonian}) and expanding the resulting trace-log to quartic order yields the magnetic free energy
\begin{eqnarray}
\label{eq:LandauFreeEnergy}
	\mathcal{F} &=& \sum_{\mbf{q}}\sum_{abcd}\left( (U^{-1})^{abcd} - \chi^{abcd}_0(\mbf{q}) \right)\mbf{M}^{ab}(\mbf{q})\mbf{M}^{cd}(-\mbf{q}) \nonumber \\
	&+& \sum_{\substack{abcd \\ efgh}} \Big[ \beta^{abcdefgh}_1 \big( \mbf{M}_X^{ab} \cdot \mbf{M}_X^{cd} \big) \big( \mbf{M}_X^{ef} \cdot \mbf{M}_X^{gh} \big) \nonumber \\
	 && \ \ \ \ \ + \beta^{abcdefgh}_2 \big( \mbf{M}_Y^{ab} \cdot \mbf{M}_Y^{cd} \big) \big( \mbf{M}_Y^{ef} \cdot \mbf{M}_Y^{gh} \big) \nonumber \\
	 && \ \ \ \ \ + \gamma^{abcdefgh} \big( \mbf{M}_X^{ab} \cdot \mbf{M}_X^{cd} \big) \big( \mbf{M}_Y^{ef} \cdot \mbf{M}_Y^{gh} \big) \nonumber \\
	 && \ \ \ \ \ + \omega^{abcdefgh} \big( \mbf{M}_X^{ab} \cdot \mbf{M}_Y^{cd} \big) \big( \mbf{M}_X^{ef} \cdot \mbf{M}_Y^{gh} \big) \Big]\,,
\end{eqnarray}
reflecting the fact that the order parameters have an orbital structure as well. Here we have maintained the $\mbf{q}$-dependence of the magnetic order parameter at quadratic level so to distinguish commensurate and incommensurate magnetic phases. The coefficient $\chi^{abcd}_0(\mbf{q})$ is the static bare spin-spin susceptibility
\begin{eqnarray}
	\chi_0^{abcd}(\mbf{q}) &=& \sum_{nm\mbf{k}} u^{a}_{n}(\mbf{k}+\mbf{q}) u^{b}_{n}(\mbf{k}+\mbf{q})^{\ast} u^{c}_{m}(\mbf{k}) u^{d}_{m}(\mbf{k})^{\ast} \nonumber \\ && \times \frac{n_F(\xi^{m}_{\mbf{k}})-n_F(\xi^{n}_{\mbf{k+q}})}{\xi^{n}_{\mbf{k+q}}-\xi^{m}_{\mbf{k}}}\,,
\end{eqnarray}
where $u^{a}_{n}(\mbf{k})$ are unitary transformations between band and orbital space and $m$ and $n$ label the bands. Here we discuss a method to evaluate the quartic coefficients appearing in the Landau free energy Eq.~(\ref{eq:LandauFreeEnergy}) by writing the magnetic order parameters in terms of the IRs of the point group and identifying which of these is the first to condense. This allows us to project the coefficients onto the appropriate IR. The decomposition in terms of IRs has the added benefit of providing an understanding of why specific orbital combinations appear together in the orbitally resolved magnetic order pa\-ra\-me\-ter.

The appropriate space groups in the paramagnetic tetragonal phase for describing the FeSCs are ${\rm P4/nmm}$ or ${\rm I4/mmm}$ and hence the relevant point group is D$_{\rm 4h}$. In the following we do not consider complications arising from nematic or strutural transitions above the magnetic transition. 
Assuming a negligible SOC implies that the related operations act separately on wavevector and orbital spaces, while spin space remains inert. The effects of SOC were considered in Ref.~\onlinecite{cvetkovic13} although in a simplified three-orbital model. In the present case we find that the magnetic order parameters $\mbf{M}_I$ belon\-ging to the relevant IRs $I=\{A_{1g},B_{1g},A_{2g},B_{2g}\}$, split into the following two- and one-dimensional IRs, $\{{\rm M}_{I,x},{\rm M}_{I,y}\}\sim I\times E_g$ and ${\rm M}_{I,z}\sim I\times A_{2g}$, respectively. 

\begin{figure}
\centering
\includegraphics[width=0.7\columnwidth]{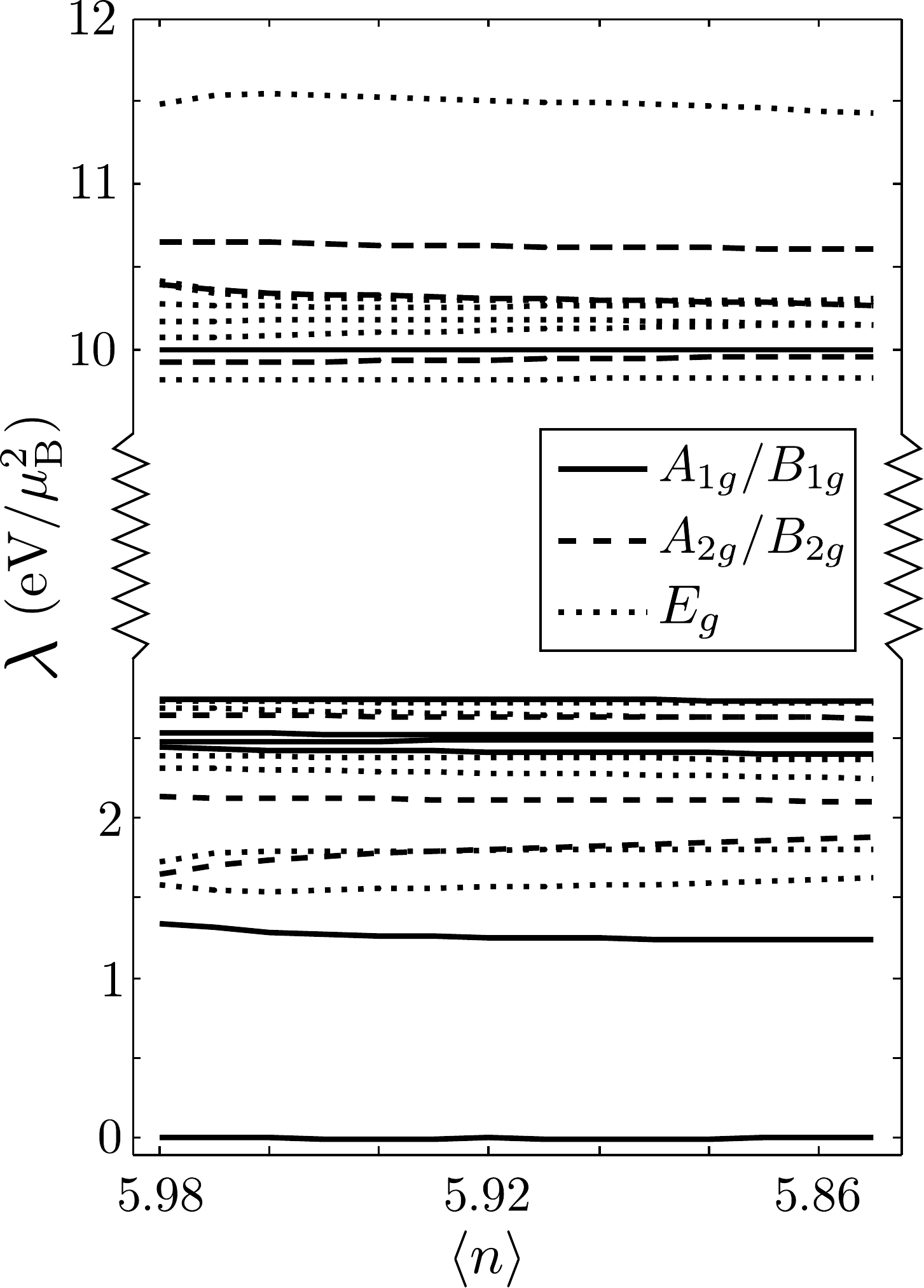}
\caption{\label{fig:eigenvalues} Eigenvalues of the inverse magnetic propagator as a function of electron filling at the transition temperature $T_{\rm SDW}$. A zero eigenvalue indicates a magnetic transition. Two distinct features are clear: The leading instability is well seperated from subleading instabilities. The eigenvalues cluster into two groups, the group nea\-rest $\lambda=0$ consists of instabilities leading to order parameters with real matrix elements in orbital space. In contrast the group around $\lambda=10$ is characterized by imaginary matrix elements for the order parameters.}
\end{figure}

The ordering wavevectors, $\mbf{Q}_X$ and $\mbf{Q}_Y$, are located at the BZ boun\-dary and satisfy $\mbf{Q}_{X,Y}=-\mbf{Q}_{X,Y}$ up to a reciprocal lattice vector. The latter property implies that the relevant ${\rm D_ {4h}}$ IRs for the present study consist of the three subsets $\{A_{1g}, B_{1g}\}$, $\{A_{2g},B_{2g}\}$ and $E_{g}$. At quadratic level the IRs of a given subset are de\-ge\-ne\-ra\-te, thus reflecting the inability of the quadratic term to distinguish between single-$\mbf{Q}$ and double-$\mbf{Q}$ phases. We stress that the form of the order parameters presented in Eq.~(\ref{eq:A1g_B1g_structure})--(\ref{eq:Eg_structure}) is not a consequence of the bandstructure considered here but follows from the D$_{\rm 4h}$ point group symmetry. The magnetic order parameters that transform according to the $\{A_{1g},B_{1g}\}$ IRs are given by the non-zero matrix elements in orbital space (see Appendix~\ref{app:irreps} for the detailed expressions):
\begin{eqnarray}
	&& \widehat{\mbf{M}}_{A_{1g}/B_{1g}} = \nonumber\\  && \begin{psmallmatrix}
			M^{xz,xz} & 0 & 0 & 0 & 0  \\
			0 & M^{yz,yz} & 0 & 0 & 0 \\
			0 & 0 &  M^{xy,xy} & 0 & 0 \\
			0 & 0 & 0 &  M^{x^2-y^2,x^2-y^2}  &  M^{x^2-y^2,z^2} \\
			0 & 0 & 0 &  M^{z^2,x^2-y^2}  &  M^{z^2,z^2}
\end{psmallmatrix}\,,\label{eq:A1g_B1g_structure}
\end{eqnarray}
while the $\{A_{2g},B_{2g}\}$ orbital composition reads:
\begin{eqnarray}
	&& \widehat{\mbf{M}}_{A_{2g}/B_{2g}} = \nonumber\\  && \begin{psmallmatrix}
			0 & M^{xz,yz} & 0 & 0 & 0 \\
			M^{yz,xz} & 0 & 0 & 0 & 0 \\
			0 & 0 & 0 & M^{xy,x^2-y^2} & M^{xy,z^2} \\
			0 & 0 & M^{x^2-y^2,xy} & 0 & 0 \\
			0 & 0 & M^{z^2,xy} & 0 & 0
\end{psmallmatrix}\,,\label{eq:A2g_B2g_structure}
\end{eqnarray}
and finally the $E_g$ corresponds to:
\begin{eqnarray}
	&& \widehat{\mbf{M}}_{E_g} = \nonumber\\ && \begin{psmallmatrix}
	0 & 0 & M^{xz,xy} & M^{xz,x^2-y^2} & M^{xz,z^2} \\
	0 & 0 & M^{yz,xy} & M^{yz,x^2-y^2} & M^{yz,z^2} \\
	M^{xy,xz} & M^{xy,yz} & 0 & 0 & 0 \\
	M^{x^2-y^2,xz} & M^{x^2-y^2,yz} & 0 & 0 & 0 \\
	M^{z^2,xz} & M^{z^2,yz} & 0 & 0 & 0
	\end{psmallmatrix}\label{eq:Eg_structure}\,.
\end{eqnarray}
As these different types of magnetic order parameters belong to distinct IRs, they are mutually exclusive. This implies that certain orbital combinations never appear in tandem, at least not as a result of a transition from a paramagnetic phase. Note that this does not rule out accidental degeneracies between the three, although for the cases studied here we find that the condensed order parameter is well separated from the others, as shown in Fig.~\ref{fig:eigenvalues}.

As stated previously, our goal is to project the quartic coefficients onto the leading magnetic instability occuring at $T_{\mathrm{SDW}}$, which can be achieved in any representation. Hence, in the orbital space picture we write the magnetic order parameters as
\begin{eqnarray}
	\mbf{M}^{ab}_{X,Y}=\mbf{M}_{X,Y}v^{ab}_{X,Y}\,.
\end{eqnarray}
The internal structure of the order parameters at the onset of the instability, reflected in the orbital space matrix $\hat{v}_{X,Y}$, can be found from the quadratic term in the free energy. The latter defines the static part of the inverse pro\-pa\-ga\-tor of the magnetic order parameter fluctuations $\mbf{M}_{X,Y}$, defined as
\begin{eqnarray}
	\left({\cal D}_{\rm mag}^{-1}(\mbf{Q}_{X,Y})\right)^{abcd}= (U^{-1})^{abcd} - \chi^{abcd}_0(\mbf{Q}_{X,Y})\,.
\end{eqnarray}
For the remaining analysis we will refer to the above simply as the inverse magnetic propagator, note however, that it is not identical to the random phase approximation (RPA) spin susceptibility.

We evaluate the inverse magnetic propagator as a function of temperature until a magnetic transition, signified by its smallest eigenvalue $\lambda^{(1)}$ crossing zero, occurs:
\begin{eqnarray}
	\left({\cal D}^{-1}_{\rm mag}(\mbf{Q}_{X,Y}) \right)^{abcd} v^{cd}_{X,Y} = \lambda^{(1)} v^{ab}_{X,Y}\,.\label{eq:instability_eq}
\end{eqnarray}
Note that this is equivalent to the divergence of the lea\-ding eigenvalue of the RPA susceptibility~\cite{MHCnem}.

The calculation of the magnetic propagator, that is a rank-4 tensor in orbital space, greatly simplifies when transferring to the IR formalism. In the latter case, one can define a single rank-2 magnetic propagator per IR ($I$), i.e. $\big({\cal D}^{-1}_{{\rm mag};I}\big)^{ss'}$. Thus the magnetic propagator can only mix different order parameters ($s,s'$), belonging to the same IR. Moreover, invariance under translations implies that the eigenvalues of $\big({\cal D}^{-1}_{{\rm mag};A_{1g}}\big)^{ss'}$ and $\big({\cal D}^{-1}_{{\rm mag};B_{1g}}\big)^{ss'}$ are degenerate, and similarly for the $\{A_{2g},B_{2g}\}$ subset, see Appendix~\ref{app:irreps}. In Fig.~\ref{fig:eigenvalues} we show the resulting eigen\-va\-lues as a function of filling. Inte\-re\-stin\-gly, for the cases explored in this paper, the smallest pair of eigenvalues always originate from the $\{A_{1g},B_{1g}\}$ subset of IRs. In fact, also the subleading pair belongs to the latter subset as shown in Fig.~\ref{fig:eigenvalues}.

The condensation of the $\{A_{1g},B_{1g}\}$ representations is consistent with Hartree-Fock methods, which find magnetic order parameters with the orbital structure given in Eq.~(\ref{eq:A1g_B1g_structure})~\cite{gastiasoro15,scherer16}, with the off-diagonal component being smaller than the diagonal components by an order of magnitude. As shown in Fig.~\ref{fig:eigenvalues}, the eigenvalues of the inverse magnetic propagator cluster into two groups. The group closest to $\lambda=0$ corresponds to the real representations of the magnetic order parameters, while the group around $\lambda=10$ corresponds to the imaginary ones.

\begin{figure}
\centering
\includegraphics[width=0.85\columnwidth]{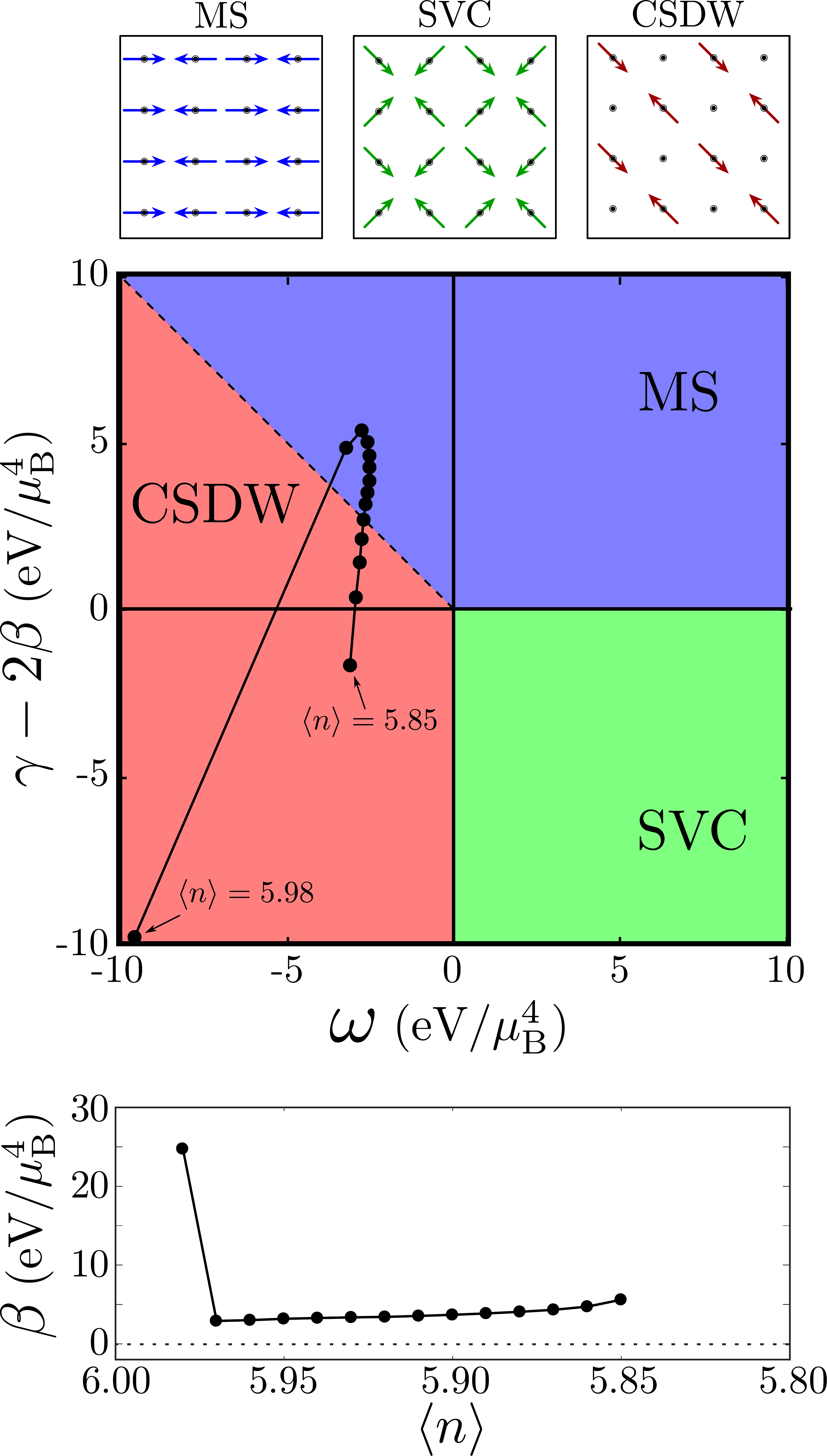}
\caption{\label{fig:commensurate_phase_diagram} (Color online) Leading instabilities for the free energy in Eq.~(\ref{eq:free_energy_simple}). The magnetic structures obtained are shown in the top panel, note that the C$_4$ orders have $|\mbf{M}_X|=|\mbf{M}_Y|$ as required by the free ener\-gy. The dots signify the microscopically obtained values of the quartic coefficients for $U=0.85\ \mathrm{eV}$ and $J=U/4$ as a function of filling for a step size of $\delta \langle n \rangle=0.01$. The evolution of the third independent quartic coefficient, $\beta$, is shown in the bottom panel as a function of doping. Finally we note that in the absence of SOC, only the relative orientation of $\mbf{M}_X$ and $\mbf{M}_Y$ is fixed by the free energy.}
\end{figure}

Having established the orbital structure of the leading instability we can proceed with evaluating the quartic coefficients. These are rank-8 tensors (see Appendix B for the explicit expressions) in orbital space, and to determine the symmetry of the magnetic order at the in\-sta\-bi\-li\-ty we project these onto the leading insta\-bi\-li\-ty using the orbital content obtained from the diagonalization of the (inverse) magnetic propagator at the insta\-bi\-li\-ty, Eq.~(\ref{eq:instability_eq}). At the instability we can then define
\begin{eqnarray}
	\beta \equiv \sum_{\substack{abcd \\ efgh}}\beta^{abcdefgh}_1 v_X^{ab}v_X^{cd}v_X^{ef}v_X^{gh}\,,\label{eq:beta1_cont} \\
	\beta \equiv \sum_{\substack{abcd \\ efgh}}\beta^{abcdefgh}_2 v_Y^{ab}v_Y^{cd}v_Y^{ef}v_Y^{gh}\,,\label{eq:beta2_cont} \\
	\gamma \equiv \sum_{\substack{abcd \\ efgh}}\gamma^{abcdefgh} v_X^{ab}v_X^{cd}v_Y^{ef}v_Y^{gh}\,, \\
	\omega \equiv \sum_{\substack{abcd \\ efgh}}\omega^{abcdefgh} v_X^{ab}v_Y^{cd}v_X^{ef}v_Y^{gh}\,, \label{eq:omega_cont}
\end{eqnarray}
and we note that while the results of the contractions in Eq.~(\ref{eq:beta1_cont}) and (\ref{eq:beta2_cont}) are identical by symmetry, the orbitally resolved coefficients $\beta^{abcdefgh}_1$ and $\beta^{abcdefgh}_2$ are related by C$_4$ rotations since these act non-trivially on orbital indices. The quartic terms in the free energy thus read
\begin{eqnarray}
	\mathcal{F}^{(4)} &=& \beta\left(\mbf{M}_X^2 + \mbf{M}_Y^2 \right)^2 + (\gamma-2\beta) \mbf{M}_X^2 \mbf{M}_Y^2 \nonumber \\  &+& \omega \left(\mbf{M}_X \cdot \mbf{M}_Y\right)^2\,. \label{eq:free_energy_simple}
\end{eqnarray}
In Fig.~\ref{fig:commensurate_phase_diagram} we show the leading instabilities associated with this free energy along with values for the quartic coefficients obtained using the method described above. With this method we can assess the effect of the orbital weights in Eqs.~(\ref{eq:beta1_cont})--(\ref{eq:omega_cont}) and thus infer the contrasts between cases where the orbital content is included or neglected.

We end this section by noting that, as is well-known, mean-field theories neglect of the effects of order pa\-ra\-me\-ter fluctuations. In the context of the pnictides, magnetic fluctuations lead to a number of preemptive orders, such as the prevalent spin-nematic order~\cite{kim11,rotundu11,prokes11}, along with the less familiar charge density wave associated with the CSDW phase, and the spin-vortex density wave associated with the SVC phase~\cite{fernandes16}. However, as these preemptive orders are decided by the same free energy coefficients as the magnetic orders, the phase boundaries between the va\-rious magnetic phases should be stable against their pre\-sen\-ce.

\section{results}

\subsection{Effects of orbital content}

\begin{figure}[t]
\centering
\includegraphics[width=0.9\columnwidth]{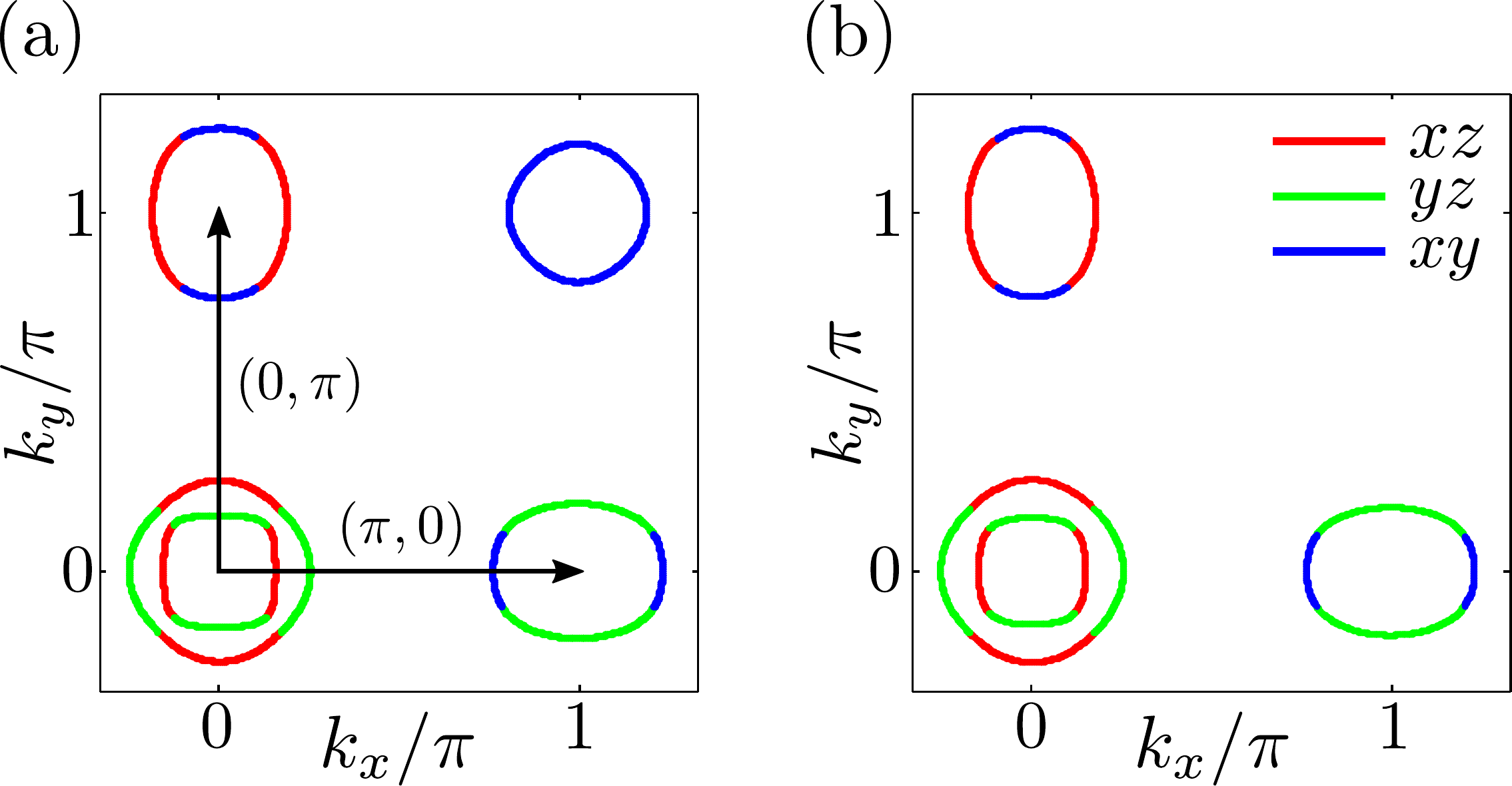}
\caption{\label{fig:Fermi_surface} (Color online) (a) Fermi surface for $\langle n \rangle =5.95$ in the paramagnetic phase. Magnetism is driven by nesting between the hole poc\-kets at $\Gamma=(0,0)$ and $\mathrm{M}=(\pi,\pi)$, and the electron pockets at $\mathrm{X}=(\pi,0)$ and $\mathrm{Y}=(0,\pi)$. (b) Fermi surface for $\langle n \rangle =6$ where the ${\rm M}$ pocket has been removed as described in the text.}
\end{figure}

To demonstrate the effects due to the presence of the orbital content we consider a specific choice of interaction parameters, $U=0.85$~eV and $J=U/4$, and evaluate the coefficients $\beta$, $\gamma$, and $\omega$ using the method described in the previous section, and the associated Appendices. At a filling of $\langle n \rangle=5.95$, for which the Fermi surface is shown in Fig.~\ref{fig:Fermi_surface}(a), we find that the magnetic order parameter is dominated by the $xy$ orbital along with the $yz$ ($xz$) orbital for magnetic orders mo\-du\-la\-ted along ${\rm X}$ (${\rm Y}$).
\begin{figure}[t]
\centering
\includegraphics[width=0.95\columnwidth]{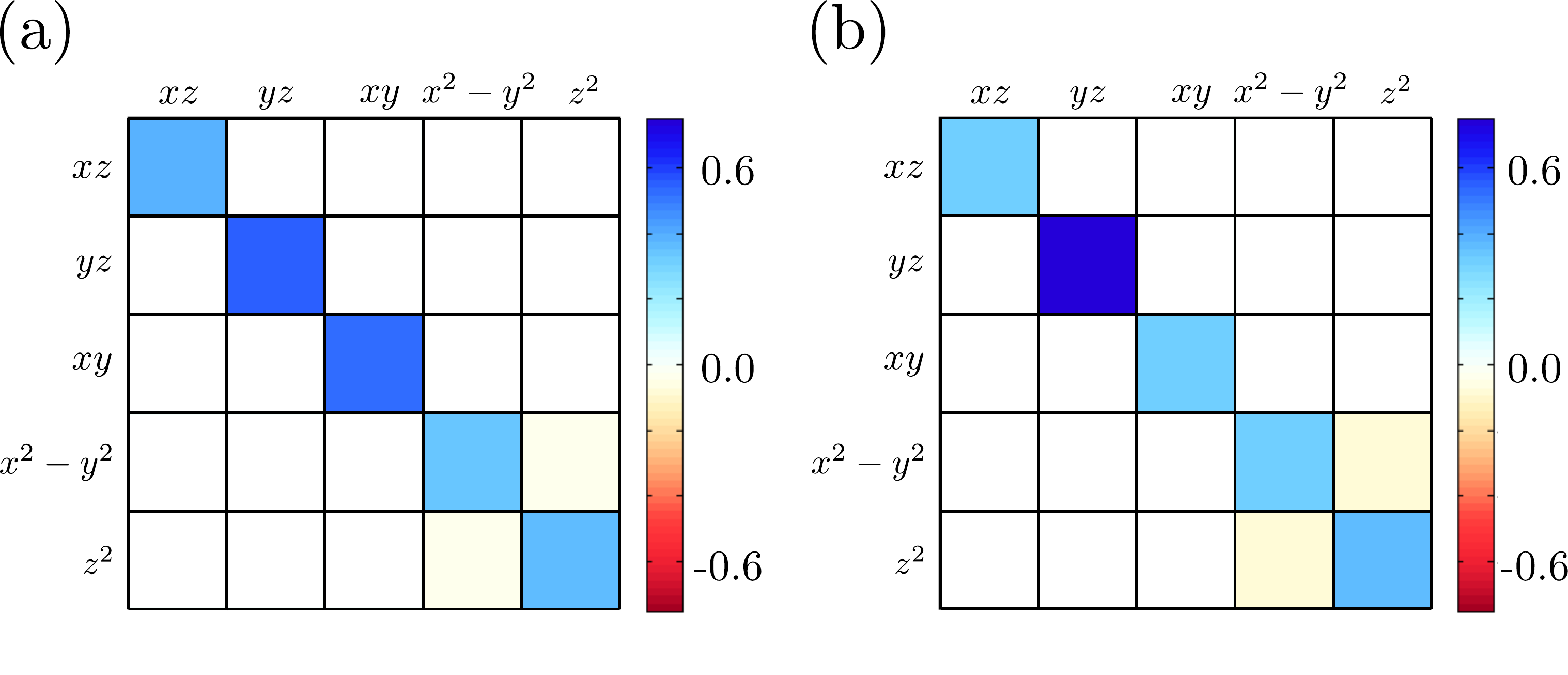}
\caption{\label{fig:orbital_content} (Color online) (a) Orbital weight $v^{ab}_X$ at the magnetic transition for the FS shown in Fig.~\ref{fig:Fermi_surface}(a), with $U=0.85\ \mathrm{eV}$, $J=U/4$, and $\langle n \rangle = 5.95$, yielding an MS phase. Note that the off-diagonal elements between $x^2-y^2$ and $z^2$ are small but non-zero. (b) Orbital weight $v^{ab}_X$ at the magnetic transition for the FS shown in Fig.~\ref{fig:Fermi_surface}(b), with $U=1.2\ \mathrm{eV}$, $J=U/6$ and $\langle n \rangle=6$ where the ${\rm M}$ pocket has been removed, hence resulting in a CSDW phase.}
\end{figure}
This is reflected in Fig.~\ref{fig:orbital_content}(a). The large $xy$ contribution arises due to nesting between the ${\rm X}$ and ${\rm Y}$ pockets and the ${\rm M}$ pocket while the $xz/yz$ contribution is due to nesting between the $\Gamma$ pocket and the X and Y pockets [cf. the Fermi surface in Fig.~\ref{fig:Fermi_surface}(a)].

\begin{figure*}
\centering
\includegraphics[width=\textwidth]{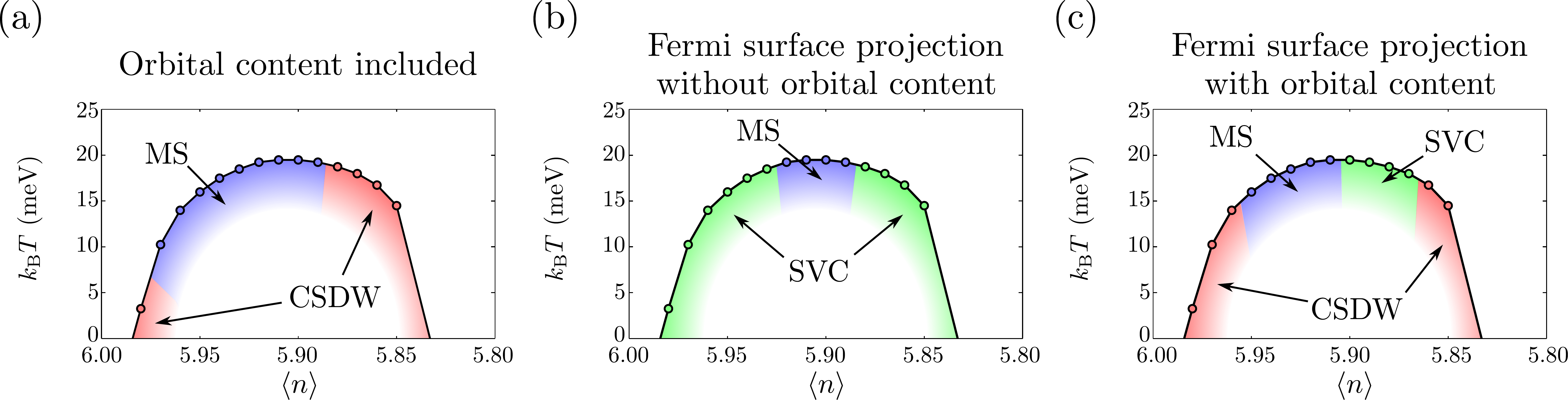}
\caption{\label{fig:phase_diagrams_U085} (Color online) Leading instabilities for the bandstructure of Ref.~\cite{ikeda10} with $U=0.85\ \mathrm{eV}$ and $J=U/4$. The color corresponds to the type of magnetic order parameter at the transition and is the same as the one in Fig.~\ref{fig:commensurate_phase_diagram}, red is the CSDW phase, blue is the magnetic stripe phase, and green is the SVC phase. As the free energy analysis with only quartic coefficients is only valid in the vicinity of the phase transition, the region of the phase diagram deep in the magnetic phase is intentionally left blank.}
\end{figure*}

In Fig.~\ref{fig:phase_diagrams_U085}(a) we show the type of the established magnetic phase near $T_{\mathrm{SDW}}$ upon varying doping and tem\-pe\-ra\-tu\-re, as obtained from the analysis of the free ener\-gy coefficients, as shown in Fig~\ref{fig:commensurate_phase_diagram}. All magnetic transitions were found to be of second order, in agreement with Ref.~\onlinecite{gastiasoro15}. As usual, the mean-field approach allows for a finite $T_{\mathrm{SDW}}$ even though it is the result of a two-dimensional calculation and overestimates the actual magnetic transition temperature. Including fluctuations in a three-dimensional calculation, i.e., taking into account the layered structure of the material will yield a finite but downward-renormalized $T_{\mathrm{SDW}}$.
As seen from Fig.~\ref{fig:phase_diagrams_U085}(a), we find a magnetic dome do\-mi\-na\-ted by the MS phase, while the CSDW phase appears on both the electron- and the hole-doped sides. As discussed previously, standard DFT-generated tight-binding bands do not generally lead to a magnetic dome centered with maximum $T_{\mathrm{SDW}}$  at $\langle n \rangle = 6.0$~\cite{gastiasoro15,timm11}.
The findings in Fig.~\ref{fig:phase_diagrams_U085}(a) are confirmed by selfconsistent Hartree-Fock calculations. The latter are performed with a momentum-space grid with $100\times100$ $\mbf{k}$-points in the 1-Fe BZ, while the quartic coefficients are computed using a grid size of $400\times400$. Overall our findings are consistent with previous stu\-dies~\cite{gastiasoro15}, though we cannot confirm the presence of the SVC phase for $U = 0.85\,$eV at any filling even down to low temperatures. We attribute this discrepancy to the slow convergence of the mean-fields. In our present study the iterative solution of the mean-field equations was stopped when the rate of change between iterations, $\epsilon$, for the dimensionless order parameters dropped below a certain value. Only for a stopping criterion $ \epsilon < 10^{-6} $ we reach the reported CSDW phase. Comparing the free energy evaluated for converged Hartree-Fock solutions between SVC and CSDW confirms that the CSDW state indeed is the preferred state among the commensurate SDW states in this parameter regime. A less strict self-consistency requirement indeed yields an SVC phase in the electron-doped region, as reported previously~\cite{gastiasoro15}.

\begin{figure}
\centering
\includegraphics[width=0.5\columnwidth]{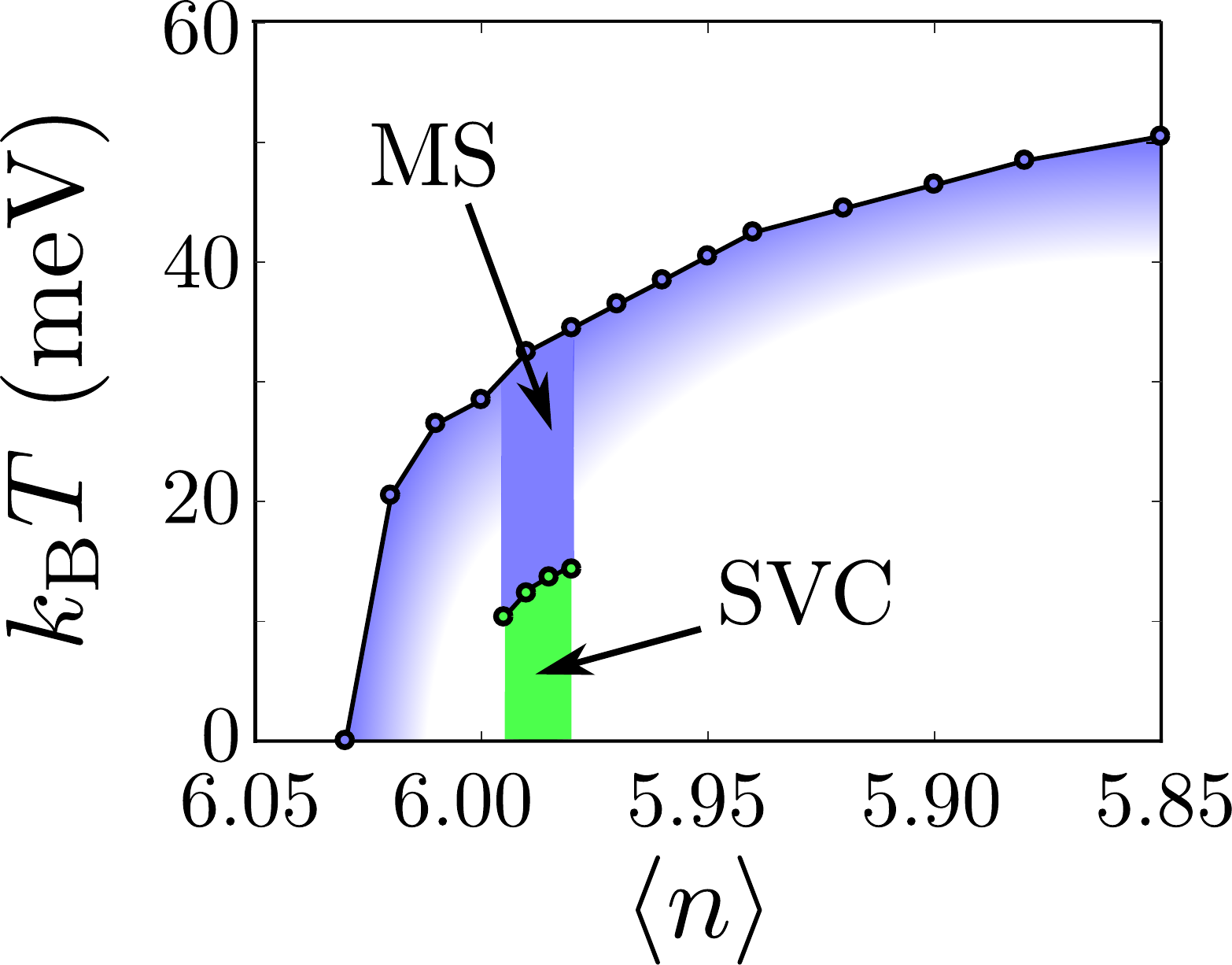}
\caption{\label{fig:U095_phase_diagram} (Color online) Leading instabilities for $U=0.95$ eV and $J=U/4$ supplemented by Hartree-Fock calculations for specific doping values confirming the existence of a secondary transition to an SVC phase at lower temperatures.}
\end{figure}

We note, however, that we find both the CSDW and the SVC phases when the interaction strength is increased to $U = 0.95\,$eV, Fig.~\ref{fig:U095_phase_diagram}, in agreement with Ref.~\onlinecite{gastiasoro15}. 
More specifically, both our analysis of the Landau free energy and our selfconsistent Hartree-Fock calculations show that for  $U = 0.95\,$eV the MS phase is realized close to $T_{\mathrm{SDW}}$ throughout the magnetic dome. Within Hartree-Fock, we confirm a subsequent change of the magnetic phase in the electron doped region, from the C$_{2}$ MS to the C$_{4}$ SVC, as the tem\-pe\-ra\-tu\-re is lowered. In agreement with Ref.~\onlinecite{gastiasoro15}, we find the transitions between magnetic states to be of second order. Note that, as shown in Ref.~\onlinecite{fernandes12}, including the effect of magnetic fluctuations drives the transitions first order. The quartic coefficients were also computed in Ref.~\onlinecite{giovannetti} using a slightly different approach, and we find broad agreement with the results presented there, albeit the interactions applied in Ref.~\onlinecite{giovannetti} are larger than the ones used in this work. In Ref.~\onlinecite{giovannetti} a large MS region was found, with an SVC region appearing upon electron doping. The energy landscape was reported to be rather flat on the electron doped side, consistent with the fact that our Hartree-Fock calculations require a high accuracy to properly converge. In contrast to our results however, first order transitions are reported in Ref.~\onlinecite{giovannetti} for a narrow doping range. This discrepancy could, however, be caused by the different methods applied.

As mentioned above, a direct PM-SVC transition was not observed for any of the parameters considered. This is in contrast to results from three-band models, which predict the dominant tetragonal phase to be the SVC phase~\cite{wang15,hoyer16a}. To shed light on this discre\-pan\-cy we eva\-lua\-te the quartic coefficients in the absence of orbital content and project onto the Fermi surface. This is achieved by fixing the matrix elements in the Green functions to unity and only including energies within a window of $50$ meV about the Fermi level. In this case we find a PM-SVC transition for a large doping range, as evidenced in Fig.~\ref{fig:phase_diagrams_U085}(b). The MS phase shrinks and the CSDW phase is entirely replaced by the SVC phase, thus connecting to the results obtained from earlier band models.

Hence the orbital content of the FS plays an important role in promoting the CSDW double-$\mbf{Q}$ phase over the SVC double-$\mbf{Q}$ phase. To illustrate the effect of the orbital content we also consider the case in which the coefficients are projected onto the FS, while the orbital content is fully retained in the Green functions. The result is shown in Fig.~\ref{fig:phase_diagrams_U085}(c) and exhibits some agreement with the case where the entire bandstructure is included. While the MS phase is still present, the C$_4$ phases are more prevalent when only low-energy contributions to the quartic coefficients are included. We interpret the fact that Figs.~\ref{fig:phase_diagrams_U085}(a) and \ref{fig:phase_diagrams_U085}(c) are not identical as evidence that both temperature effects and finite energy particle-hole excitations are important for determining the symmetry of the magnetic order parameter~\cite{kreisel}.

\subsection{Impact of the M-pocket}

\begin{figure}
\centering
\includegraphics[width=\columnwidth]{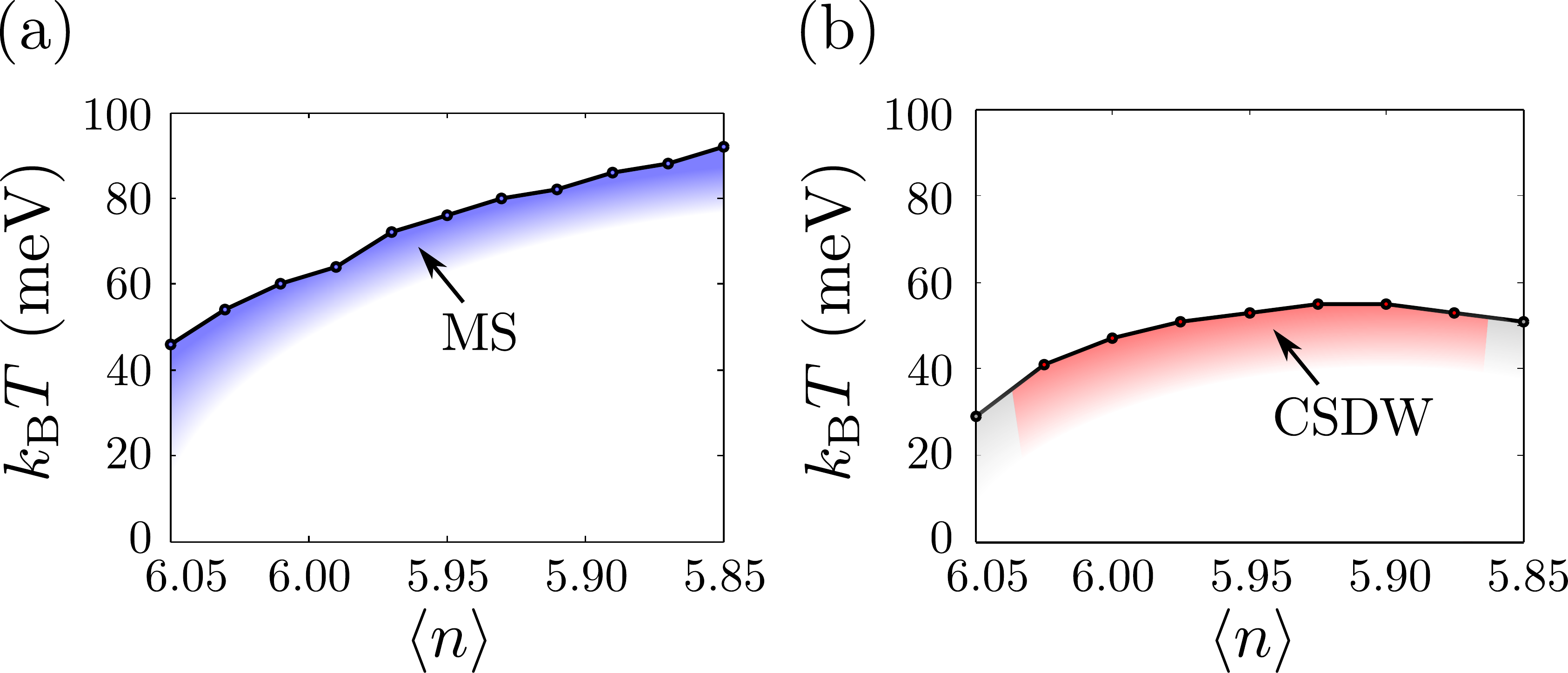}
\caption{\label{fig:m_pocket_phase_diagram} (Color online) Phase diagrams for the parameters $U=1.2\ \mathrm{eV}$ and $J=U/6$. (a) Leading instability for the original bandstructure is shown and the system exhibits an MS phase throughout. (b) Leading instability in the case of an absent hole pocket at M. The MS phase of the original case is replaced with a CSDW phase, while the transition temperature is reduced due to the change in nesting properties of the Fermi surface. The grey area denotes regions of incommensurate magnetism.}
\end{figure}

The strong influence of the $xy$ orbital along with the fact that it is mainly encountered in the ${\rm M}$ pocket, which only appears in certain materials, prompt additional calculations. Below we compare calculations carried out with or without the hole-pocket at M. For the current bandstructure, the ${\rm M}$-pocket can be removed by adding to the Hamiltonian a C$_4$ invariant term $\epsilon_{ab}\left( 1 + \cos k_x \cos k_y \right)$ where $\epsilon_{ab}=-0.1\,\mathrm{eV}\delta_{a,xy}\delta_{b,xy}$. In order to avoid magnetic transition temperatures that become too small to resolve numerically, we switch to study a case with interaction para\-me\-ters given by $U=1.2\ \mathrm{eV}$ and $J=U/6$. For the original bandstructure this reproduces an MS phase in the region of the phase diagram examined here, as shown in Fig.~\ref{fig:m_pocket_phase_diagram}(a). However, as the ${\rm M}$-pocket is removed, the single-$\mbf{Q}$ MS phase is replaced by the double-$\mbf{Q}$ CSDW phase, as seen in Fig.~\ref{fig:m_pocket_phase_diagram}(b). This finding is confirmed by Hartree-Fock calculations, which additionally finds that in this case the two double-$\mbf{Q}$ phases are nearly degenerate.

The above results can be understood as follows. First one notes that when the M pocket is present, it facilitates the simultaneous nesting of all the pockets, with a single wavevector, thus promoting the MS phase. The latter argument is supported by the findings presented in Fig.~\ref{fig:orbital_content}(a), where we find that the $yz$ ($xz$) and the $xy$ contributes roughly equally to the nesting along $\mbf{Q}_X$ ($\mbf{Q}_Y$). In contrast, when the ${\rm M}$ pocket is absent the formation of a MS phase is expected to be suppressed, since single-$\mbf{Q}$ nesting will unavoidably leave the FS of one electron pocket ungapped. As the latter situation is in general energetically costly, the system can instead prefer to stabilize a double-$\mbf{Q}$ phase, via nesting both the $yz$ orbital part of the ${\rm X}$ pocket and the $xz$ orbital part of the ${\rm Y}$ pocket with the inner hole pocket at ${\rm \Gamma}$. In this case, nesting along $\mbf{Q}_X$ ($\mbf{Q_Y}$) is dominated by $yz$ ($xz$), as it is found in Fig.~\ref{fig:orbital_content}(b).

At the level of the quartic coefficients, the value of $\gamma - 2\beta$ is lowered upon the removal of the M-pocket, thus favoring the double-$\mbf{Q}$ C$_4$-symmetric phases. Furthermore, the value of $\omega$ is observed to be very close to zero, reflecting the near-degeneracy observed in Hartree-Fock calculations. It is tempting to associate these fin\-dings with the modification of the FS nesting properties, directly discernable from comparing Figs.~\ref{fig:Fermi_surface}(a) and \ref{fig:Fermi_surface}(b). However, Hartree-Fock calculations show that the MS phase reappears at lower temperatures and higher values of $U$, leading us to conclude that, while the absence of the hole-pocket at M disfavors the C$_2$ MS phase, it does not rule it out.

\section{Conclusions}

In this work we focused on the role of the orbital degrees of freedom in the formation of commensurate magnetism for a realistic itinerant model of FeSC. Throughout our analysis we classified the orbitally resolved magnetic order parameters in terms of the irreducible representations of the point group ${\rm D_{4h}}$. Employing this formalism greatly facilitated the multiorbital analysis, and allowed us to show that the structure of the magnetic order parameters fall into three distinct categories, spanned by the IR subsets $\{A_{1g},B_{1g}\}$, $\{A_{2g},B_{2g}\}$ and $E_{g}$. 

For the parameter values studied in this paper we obtained that the $\{A_{1g},B_{1g}\}$ IRs condense first, corresponding to a nearly-diagonal magnetic order parameter in orbital space. This finding is consistent with pre\-vious results retrieved via Hartree-Fock calculations~\cite{gastiasoro15,scherer16}. On the other hand, magnetic order parameters with non-zero off-diagonal elements in orbital space might appear as a consequence of increasing the ratio $U'/U$. The latter can enhance the inter-orbital contributions in Eq.~(\ref{eq:mag_mean_field_eq}) and promote the $\{A_{2g},B_{2g}\}$ or $E_g$ IRs. However, from the constraints imposed by SO(3) rotational invariance~\cite{oles83}, an increase in $U'$ entails a decrease in Hund's coupling, which however is also believed to be substantial~\cite{devereaux09,shen09,kotliar11}. Interestingly, recent renormalization group (RG) stu\-dies~\cite{chubukov16} indicate that $U=U'$ is a fixed trajectory of the RG flow even for finite Hund's coupling thus vio\-la\-ting the assumed rotational invariance of the microscopic interactions~\cite{rot_inv}. 

Besides our generic conclusions regarding the most prominent IR of the stabilized magnetic order parameter, in this work we further investigated the effect of the bandstructure's orbital character on the order's type, i.e. MS, CSDW or SVC, via contrasting various ways of including the orbital content. We found that in the absence of orbitals, the preferred C$_4$ magnetic structure is the SVC phase, in agreement with calculations based on three-band mo\-dels~\cite{wang15,hoyer16a}. With the inclusion of orbital content, however, the SVC phase gives way to the CSDW phase, whose magnetic structure is consistent with the experi\-men\-tal\-ly observed magnetic C$_4$ phase~\cite{allred16a}. 

Additionally, we studied the impact of the ${\rm M}$ pocket on the resulting symmetry of the order parameter of the magnetic ground state. In the presence of a hole pocket at ${\rm M}$ we find a C$_2$ magnetic phase driven primarily by intraorbital $xy/xy$ and $yz (xz)/yz(xz)$ magnetic fluctua\-tions along $\mbf{Q}_X$ ($\mbf{Q}_Y$). While we find that the M-pocket is important to the formation of a C$_2$ symmetric phase, its absence does not rule out the appearance of a C$_2$ phase.
The bandstructure considered here was independent of $k_z$, however some FeSC can show rather anisotropic dispersions along the $k_z$-direction, with the M pocket only being present for certain values of $k_z$~\cite{feng11,ding11,suzuki14}. The na\-tu\-ral next step towards a complete understanding of the effect of the M pocket on the preferred magnetic order parameter is to extend our investigation to such dispersions. 

We end with a brief remark on the fate of our results in the presence of SOC. The latter breaks the SO(3) spin-rotational invariance, and its effects can be described by adding the following quadratic term to the free energy~\cite{cvetkovic13,christensen15}
\begin{eqnarray}
\mathcal{F}_{\rm SOC}&=& \alpha_1 \left(M_{X,x}^2 + M_{Y,y}^2 \right)+\alpha_2 \left(M_{X,y}^2 + M_{Y,x}^2 \right) \nonumber \\ 
&+& \alpha_3 \left(M_{X,z}^2 + M_{Y,z}^2 \right)\,,\label{eq:soc_free_energy}
\end{eqnarray}
with $x,y,z$ denoting the spin components of the $\mbf{M}_{X,Y}$ order parameter. Explicit expressions for the $\alpha_{i}$-coefficients were given in Ref.~\onlinecite{christensen15} where the magnetic anisotropy induced by a $\lambda\mbf{L}\cdot\mbf{S}$-term was considered.
The term ${\cal F}_{\rm SOC}$ modifies the free energy at the quadratic level with respect to the order parameters and therefore plays an important role in the determination of the leading magnetic instability. Beyond the differentiation of directions in spin space, SOC may even support new types of magnetic states, constituting an interesting future avenue of research.

\begin{acknowledgments}

We thank A. V. Chubukov, R. M. Fernandes, and M. N. Gastiasoro for useful discussions. M.H.C. and B.M.A acknowledge financial support from a Lundbeckfond fellowship (Grant No. A9318).

\end{acknowledgments}

\appendix

\section{Decomposition into irreducible representations of D$_{\bm{{\rm 4h}}}$}\label{app:irreps}

In this section we present how the magnetic order parameters $\widehat{\mbf{M}}_X$ and $\widehat{\mbf{M}}_Y$ transform under the point group symmetry operations. Here the $\widehat{\phantom{M}}$ denotes matrices in orbital space. The wavevectors considered here are located at the boundary of the Brillouin zone and thus satisfy $\mbf{Q}_{X,Y} = -\mbf{Q}_{X,Y}$ up to reciprocal lattice vectors. In this section we outline the restrictions the point group symmetries impose on the orbital structure of the order parameters.

The relevant space groups for the FeSC are the P4/nmm and the I4/mmm both of which have the same point group, D$_{\rm 4h}$. Assuming negligible SOC renders the spin-indices inert under the point group transformations, and additionally implies that the non-symmorphic nature of P4/nmm does not enter the considerations.  Moreoever, since both the magnetic ordering vectors $\mbf{Q}_X$ and $\mbf{Q}_Y$ and the Fe 3$d$-orbitals are invariant under inversion $\mathcal{I}$, the symmetry classification can proceed according to the subgroup C$_{4v} \subset {\rm D_{4h}}$. The ope\-ra\-tions contained in C$_{4v}$ are the identity $E$, two $C_4(z)$ rotations, a $C_2(z)$ rotation, two $\sigma_v$ mirror operations and two $\sigma_d$ mirror operations. We note that these act se\-pa\-ra\-te\-ly on the wavevector and orbital spaces. 

For demonstrating how the point group symmetry ope\-ra\-tions act on wavevector space, let us consider for simplicity a scalar object $f$ labelled by one of the two wavevectors, i.e. $f_{\bm{Q}_X}$ and $f_{\bm{Q}_Y}$. One then finds  
\begin{eqnarray*}
	&&E \begin{pmatrix}
		f_{\mbf{Q}_X} \\
		f_{\mbf{Q}_Y}
	\end{pmatrix}
	= C_{2}(z)\begin{pmatrix}
		f_{\mbf{Q}_X} \\
		f_{\mbf{Q}_Y}
	\end{pmatrix}
	= \sigma_v \begin{pmatrix}
		f_{\mbf{Q}_X} \\
		f_{\mbf{Q}_Y}
	\end{pmatrix}
	= \begin{pmatrix}
		f_{\mbf{Q}_X} \\
		f_{\mbf{Q}_Y}
	\end{pmatrix},\\
	&&\qquad\quad C_4(z) \begin{pmatrix}
		f_{\mbf{Q}_X} \\
		f_{\mbf{Q}_Y}
	\end{pmatrix}
	= \sigma_d \begin{pmatrix}
		f_{\mbf{Q}_X} \\
		f_{\mbf{Q}_Y}
	\end{pmatrix}
	= \begin{pmatrix}
		f_{\mbf{Q}_Y} \\
		f_{\mbf{Q}_X}
	\end{pmatrix}\,.
\end{eqnarray*}
The above 2d representation is reducible and can be decomposed into the two 1d IRs, $A_{1g}$ and $B_{1g}$:
\begin{eqnarray}
	A_{1g}: \ f_{\mbf{Q}_X} + f_{\mbf{Q}_Y}, \qquad B_{1g}: \ f_{\mbf{Q}_X}-f_{\mbf{Q}_Y}\,.
\end{eqnarray}
The above analysis implies that the linear combinations $\widehat{\mbf{M}}_X \pm \widehat{\mbf{M}}_Y$ carry the 1d IRs $A_{1g}$ and $B_{1g}$, respectively.

Notably, the wavevector dependence of the order parameters additionally implies that they transform non-trivially under translations. The latter are generated by $t_{\mbf{R}}f(\mbf{r})=f(\mbf{r}+\mbf{R})$ where $\mbf{R}$ is a vector commensurate with the lattice. In momentum space we find $t_{\mbf{R}}f_{\mbf{q}} = e^{i\mbf{q}\cdot\mbf{R}}f_{\mbf{q}}$, implying that translations act as U(1) gauge transformations in momentum space. By setting the lattice constant equal to unity, any translation can be generated by a combination of unit translations along the $x$ and $y$ axes, $t_{(1,0)}$ and $t_{(0,1)}$, which act on the function $f_{\mbf{Q}_{X,Y}}$ in the following manner:
\begin{eqnarray}
	t_{(1,0)} \begin{pmatrix}
		f_{\mbf{Q}_X} \\
		f_{\mbf{Q}_Y}
	\end{pmatrix} &=&
	\begin{pmatrix}
		-1 & 0 \\
		 0 & 1
	\end{pmatrix}
	\begin{pmatrix}
		f_{\mbf{Q}_X} \\
		f_{\mbf{Q}_Y}
	\end{pmatrix}, \\
	t_{(0,1)} \begin{pmatrix}
		f_{\mbf{Q}_X} \\
		f_{\mbf{Q}_Y}
	\end{pmatrix} &=&
	\begin{pmatrix}
		 1 &  0 \\
		 0 & -1
	\end{pmatrix}
	\begin{pmatrix}
		f_{\mbf{Q}_X} \\
		f_{\mbf{Q}_Y}
	\end{pmatrix}\,,
\end{eqnarray}
and hence $t_{(1,0)/(0,1)}\left(f_{\mbf{Q}_X} + \alpha f_{\mbf{Q}_Y} \right)= \mp \left(f_{\mbf{Q}_X} - \alpha f_{\mbf{Q}_Y} \right)$ for $\alpha = \pm 1$. Translational invariance thus implies that the free e\-ner\-gy should be invariant under exchange of the IR indices $A \leftrightarrow B$. This manifests itself in the degeneracy of the eigenvalues of $\big({\cal D}^{-1}_{{\rm mag};A_{1g}/A_{2g}}\big)^{ss'}$ and $\big({\cal D}^{-1}_{{\rm mag};B_{1g}/B_{2g}}\big)^{ss'}$, as discussed in the main text.

\begin{table}[t!]
\begin{tabular}{|c||c|c|c|c|c|}
\hline
 	$\otimes$ & $A_{1g}$ & $B_{1g}$ & $A_{2g}$ & $B_{2g}$ & $E_{g}$ \\
 \hline
 \hline
 $A_{1g}$ & $A_{1g}$ & $B_{1g}$ & $A_{2g}$ & $B_{2g}$ & $E_{g}$ \\
 \hline
 $B_{1g}$ & $B_{1g}$ & $A_{1g}$ & $B_{2g}$ & $A_{2g}$ & $E_{g}$ \\
 \hline 
 $A_{2g}$ & $A_{2g}$ & $B_{2g}$ & $A_{1g}$ & $B_{1g}$ & $E_{g}$ \\
 \hline
 $B_{2g}$ & $B_{2g}$ & $A_{2g}$ & $B_{1g}$ & $A_{1g}$ & $E_{g}$ \\
 \hline
 $E_{g}$ & $E_{g}$ & $E_{g}$ & $E_{g}$ & $E_{g}$ & $A_{1g} \oplus A_{2g} \oplus B_{1g}  \oplus B_{2g}$\\
 \hline
\end{tabular}
\caption{\label{tab:decomposition_table} Decomposition of the product of two gerade D$_{\rm 4h}$ irreducible representations.}
\end{table}

On the other hand, the Fe $d$-orbitals transform accor\-ding to the following D$_{\rm 4h}$ IRs:
\begin{eqnarray}
	&& (d_{xz},d_{yz}) \sim E_g\,, \ d_{xy} \sim B_{2g}\,, \\
	&& d_{x^2-y^2} \sim B_{1g}\,, \ \ \ d_{z^2} \sim A_{1g}\,.
\end{eqnarray}
Since the magnetic order parameters constitute rank-2 tensors in the $d$-orbital space, their matrix elements can be symmetry classified according to a Kronecker product of two IRs of the point group. By employing the decomposition rule for the IRs of interest presented in Table~\ref{tab:decomposition_table} we can classify the magnetic order parameters using the IRs of D$_{\rm 4h}$. 

To exemplify the method let us focus on the $\{d_{xz},d_{yz}\}$ subspace and consider a $2\times2$ Hermitian matrix $\hat{f}$ defined in the latter space. By taking into account that the two component orbital basis in this subspace transforms according to the $E_g$ IR, the possible IRs for the $\hat{f}$ matrix are retrieved from the product decomposition
\begin{eqnarray}
	E_g \otimes E_g = A_{1g} \oplus A_{2g} \oplus B_{1g} \oplus B_{2g}\,.
\end{eqnarray}
The corresponding matrices in orbital space can be inferred from linear combinations of the functions $f_{xz,xz}$, $f_{yz,yz}$, $f_{xz,yz}$, and $f_{yz,xz}$:
\begin{eqnarray} 
A_{1g}: \ f_{xz,xz} + f_{yz,yz},\,&\qquad&  B_{1g}: \ f_{xz,xz} - f_{yz,yz},\qquad \\
A_{2g}: \ if_{xz,yz} - if_{yz,xz},\,&\qquad&  B_{2g}: \ f_{xz,yz} + f_{yz,xz}\,,\qquad
\end{eqnarray}
where e.g. $f_{xz,xz}$ is a function transforming under the point group transformations and the $i$ was included so to guaranteee the assumed Hermitian character of $\hat{f}$. The above linear combinations yield the following IR matrices:
\begin{eqnarray}
	&& A_{1g}: \ \begin{pmatrix}
		1 & 0 & 0 & 0 & 0 \\
		0 & 1 & 0 & 0 & 0 \\
		0 & 0 & 0 & 0 & 0 \\
		0 & 0 & 0 & 0 & 0 \\
		0 & 0 & 0 & 0 & 0
	\end{pmatrix}\,,\quad
	 B_{1g}: \ \begin{pmatrix}
		1 & 0 & 0 & 0 & 0 \\
		0 & -1 & 0 & 0 & 0 \\
		0 & 0 & 0 & 0 & 0 \\
		0 & 0 & 0 & 0 & 0 \\
		0 & 0 & 0 & 0 & 0
	\end{pmatrix}\,,\quad\\
	&& A_{2g}: \ \begin{pmatrix}
		0 & -i & 0 & 0 & 0 \\
		i & 0 & 0 & 0 & 0 \\
		0 & 0 & 0 & 0 & 0 \\
		0 & 0 & 0 & 0 & 0 \\
		0 & 0 & 0 & 0 & 0
	\end{pmatrix}\,,\quad
	B_{2g}: \ \begin{pmatrix}
		0 & 1 & 0 & 0 & 0 \\
		1 & 0 & 0 & 0 & 0 \\
		0 & 0 & 0 & 0 & 0 \\
		0 & 0 & 0 & 0 & 0 \\
		0 & 0 & 0 & 0 & 0
	\end{pmatrix}\,.\quad\qquad
\end{eqnarray}
Therefore, based on the above one can infer the orbital-subspace transformation properties of the magnetic order parameters. By properly extending the above analysis so to include all five orbitals one can retrieve the complete action of the symmetry operations in orbital space.   

To obtain the full transformation properties of the magnetic order parameters we combine orbital and wavevector transformation properties. As shown above, linear combinations of the wavevectors fall into either $A_{1g}$ or $B_{1g}$ IRs, while the orbitals have components in all of the four 1d IRs. Since, the leading and subleading instabilities are described by order parameters belonging to the $\{A_{1g},B_{1g}\}$ subset of IRs, we present below only the latter:
\begin{widetext}
\begin{eqnarray}
	\widehat{\mbf{M}}_{\stackrel{A_{1g},1}{B_{1g}\phantom{,1}}} \sim \begin{pmatrix}
		1 & 0 & 0 & 0 & 0 \\
		0 & 1 & 0 & 0 & 0 \\
		0 & 0 & 0 & 0 & 0 \\
		0 & 0 & 0 & 0 & 0 \\
		0 & 0 & 0 & 0 & 0		
	\end{pmatrix}_{\mbf{Q}_X \pm \mbf{Q}_Y},\,\,
	&& \widehat{\mbf{M}}_{\stackrel{A_{1g},2}{B_{1g}\phantom{,2}}} \sim \begin{pmatrix}
		1 & 0 & 0 & 0 & 0 \\
		0 & -1 & 0 & 0 & 0 \\
		0 & 0 & 0 & 0 & 0 \\
		0 & 0 & 0 & 0 & 0 \\
		0 & 0 & 0 & 0 & 0		
	\end{pmatrix}_{\mbf{Q}_X \mp \mbf{Q}_Y},\,\,
	\widehat{\mbf{M}}_{\stackrel{A_{1g},3}{B_{1g}\phantom{,3}}} \sim \begin{pmatrix}
		0 & 0 & 0 & 0 & 0 \\
		0 & 0 & 0 & 0 & 0 \\
		0 & 0 & 1 & 0 & 0 \\
		0 & 0 & 0 & 0 & 0 \\
		0 & 0 & 0 & 0 & 0		
	\end{pmatrix}_{\mbf{Q}_X \pm \mbf{Q}_Y},\quad\\
	\widehat{\mbf{M}}_{\stackrel{A_{1g},4}{B_{1g}\phantom{,4}}} \sim \begin{pmatrix}
		0 & 0 & 0 & 0 & 0 \\
		0 & 0 & 0 & 0 & 0 \\
		0 & 0 & 0 & 0 & 0 \\
		0 & 0 & 0 & 1 & 0 \\
		0 & 0 & 0 & 0 & 0		
	\end{pmatrix}_{\mbf{Q}_X \pm \mbf{Q}_Y},\,\,
	&& \widehat{\mbf{M}}_{\stackrel{A_{1g},5}{B_{1g}\phantom{,5}}} \sim \begin{pmatrix}
		0 & 0 & 0 & 0 & 0 \\
		0 & 0 & 0 & 0 & 0 \\
		0 & 0 & 0 & 0 & 0 \\
		0 & 0 & 0 & 0 & 0 \\
		0 & 0 & 0 & 0 & 1		
	\end{pmatrix}_{\mbf{Q}_X \pm \mbf{Q}_Y},\,\,
	\widehat{\mbf{M}}_{\stackrel{A_{1g},6}{B_{1g}\phantom{,6}}} \sim \begin{pmatrix}
		0 & 0 & 0 & 0 & 0 \\
		0 & 0 & 0 & 0 & 0 \\
		0 & 0 & 0 & 0 & 0 \\
		0 & 0 & 0 & 0 & 1 \\
		0 & 0 & 0 & 1 & 0		
	\end{pmatrix}_{\mbf{Q}_X \mp \mbf{Q}_Y},\quad\\
	&& \widehat{\mbf{M}}_{\stackrel{A_{1g},7}{B_{1g}\phantom{,7}}} \sim \begin{pmatrix}
		0 & 0 & 0 & 0 & 0 \\
		0 & 0 & 0 & 0 & 0 \\
		0 & 0 & 0 & 0 & 0 \\
		0 & 0 & 0 & 0 & -i \\
		0 & 0 & 0 & i & 0		
	\end{pmatrix}_{\mbf{Q}_X \mp \mbf{Q}_Y}\,,
\end{eqnarray}
\end{widetext}
where we made use of the shorthand notation $f_{\mbf{Q}_X\pm\mbf{Q}_Y}\equiv f_{\mbf{Q}_X}\pm f_{\mbf{Q}_y}$.

Combining these matrices yields the matrix order parameter given in Eq.~(\ref{eq:A1g_B1g_structure}) while the structure in Eq.~(\ref{eq:A2g_B2g_structure}) or Eq.~(\ref{eq:Eg_structure}) is found from the corresponding $\{A_{2g},B_{2g}\}$ or $E_g$ IR space defined magnetic order parameters. These matrices allow us to write the magnetic order parameter in the space of IRs:
\begin{eqnarray}
	\mbf{M}^{ab}(\mbf{Q}_i)=\sum_{I,s} \Lambda_{I,s}^{ab}(\mbf{Q}_i)\mbf{M}_{I,s}\,,\label{eq:trafoLambda}
\end{eqnarray}
with $I\in \{A_{1g},B_{1g},A_{2g},B_{2g},E_g\}$ while $s$ denotes the number of order parameters that belong to the same IR, e.g. $s=7$ for the $A_{1g}$ and similar for the $B_{1g}$ IRs. Via employing Eq.~\eqref{eq:trafoLambda} we can rewrite the bare susceptibility in the space of IRs, in which it becomes a block-diagonal matrix, with each block corresponding to a different IR. 

\begin{widetext}

\section{Quartic coefficients in orbital space}

The derivation of the quartic coefficients is straightforward but tedious. Here we present the results of carrying out the trace-log expansion to fourth order~\cite{fernandes12}. We find the coefficients to be
\begin{eqnarray}
	\beta_1^{abcdefgh} &=& \frac{1}{16}\sum_{k}\big[G^{ha}G^{bc}_X G^{de} G_X^{fg}  + G^{da}G^{bg}_X G^{he} G_X^{fc} - G^{ha}G^{be}_X G^{fc} G_X^{dg} \big] ,\\
	\beta_2^{abcdefgh} &=& \frac{1}{16}\sum_{k}\big[G^{ha}G^{bc}_Y G^{de} G_Y^{fg} + G^{da}G^{bg}_Y G^{he} G_Y^{fc} - G^{ha}G^{be}_Y G^{fc} G_Y^{dg} \big] ,\\
	\gamma^{abcdefgh} &=& \frac{1}{16}\sum_{k}\big[G^{ha} G_X^{bc} G^{de} G_Y^{fg} + G^{hc} G_X^{da} G^{be} G_Y^{fg} - G^{ha} G_X^{be} G_{X+Y}^{fc} G_Y^{dg} \nonumber \\ && \qquad \ \ - G^{hc} G_X^{de} G_{X+Y}^{fa} G_Y^{bg} + G^{fa} G_X^{bc} G^{dg} G_Y^{he} + G^{fc} G_X^{da} G^{bg} G_Y^{he} \big] ,\\
	\omega^{abcdefgh} &=& \frac{1}{8}\sum_{k}\big[G^{he} G_X^{fa} G^{bc} G_Y^{dg} + G^{ha} G^{bc}_X G^{de}_{X+Y} G^{fg}_Y - G^{ha} G_X^{be} G^{fc} G_Y^{dg} \nonumber \\ && \qquad \ \ + G^{he} G^{fc}_X G^{da}_{X+Y} G^{bg}_Y + G^{da} G^{be}_X G^{fg} G^{hc}_Y - G^{de} G^{fa}_X G^{bg} G^{hc}_Y \big]\,.
\end{eqnarray}
Here the $G^{ab}$ represent the Green functions:
\begin{eqnarray}
	G^{ab}\equiv G^{ab}(\mbf{k},i\omega_m) = \sum_{n} \frac{u^{a}_{n}(\mbf{k})u^{b}_{n}(\mbf{k})^{\ast}}{i\omega_m-E^{n}(\mbf{k})},\,\quad
	G^{ab}_{X,Y}\equiv G^{ab}(\mbf{k}+\mbf{Q}_{X,Y},i\omega_m),\,\quad
	G^{ab}_{X+Y}\equiv G^{ab}(\mbf{k}+\mbf{Q}_X+\mbf{Q}_Y,i\omega_m)\,,\qquad
\end{eqnarray}
where $i\omega_m$ is a fermionic Matsubara frequency.

\end{widetext}


\begin{thebibliography}{}
\bibitem{commensurability} We find that the inclusion of the Hartree-shifts increases the portion of commensurate versus incommensurate magnetism in the respective magnetic dome. See also \cite{christensen16}.
\bibitem{rot_inv} Note that this is not unexpected. Due to the crystal field environment the assumed rotational invariance is already broken down to the fourfold discrete subgroup, that already dictates the bandstructure. Thus the RG flow simply allows the interaction parameters to indirectly feel the effect of the crystal field.
\end{thebibliography}
\end{document}